\documentclass{iaus}
\usepackage{graphicx}

\title[Radial-velocity noise induced by magnetic cycles] %% give here short title %%
{Stellar noise and planet detection. \\
{\Large II. Radial-velocity noise induced by magnetic cycles}}

\author[X. Dumusque et al.]   %% give here short author list %%
{X. Dumusque$^{1,}$$^2$,
C. Lovis$^1$,
S. Udry$^1$
\and N. C. Santos$^2$}

\affiliation{$^1$Observatoire de Gen\`eve, 51 ch. des Maillettes, 1290 Sauverny, Switzerland \\[\affilskip]
$^2$Centro de Astrof{\'\i}sica, Universidade do Porto, Rua das Estrelas, 4150-762 Porto, Portugal}

\pubyear{2010}
\volume{276}  %% insert here IAU Symposium No.
\pagerange{0--1}
\jname{The Astrophysics of Planetary Systems: Formation, \\Structure, and Dynamical Evolution}
\editors{Alessandro Sozzetti, Mario G. Lattanzi \& Alan P. Boss, eds.}

\begin{document}

\maketitle

\begin{abstract}
For the 451 stars of the HARPS high precision program, we study correlations between the radial-velocity (RV) variation and other parameters of the Cross Correlated Function (CCF). After a careful target selection, we found a very good correlation between the slope of the RV-activity index (log(R'$_{HK}$)) correlation and the T$_{\textrm{eff}}$ for dwarf stars. This correlation allow us to correct RV from magnetic cycles given the activity index and the T$_{\textrm{eff}}$.
\\
\end{abstract}


\noindent
\underline{Introduction}

\vspace{0.1cm}
The Sun has a 11-years magnetic cycle, during which the activity level (\cite[Noyes \etal\ (1984)]{Noyes-1984}) varies between log(R'$_{HK}$)=-5 at minimum and log(R'$_{HK}$)=-4.75 at maximum. This cycle can also be seen looking at the total number of sunspots or the Sun luminosity variation.
In a recent paper, \cite[Meunier \etal\ (2010)]{Meunier-2010} show a correlation between the Sun RV variation and its magnetic cycle. Amplitudes of tenths of meter per second could be induced by such cycles, hiding signals of long period small-mass planets.

Solar type stars share the property of having an external convective enveloppe. Since upward
flows of convection (granules) have a total surface more important than downward flows
(intergranules), the stellar spectrum will be blushifted (\cite[Dravins (1982)]{Dravins-1982} and \cite[Gray (2009)]{Gray-2009}). When the activity level increases, the number of magnetic features (spots and plages) raise up, and since
these regions are known to inhibit the convection due to strong magnetic fields, the stellar spectrum will be shifted to the red. Since the activity level is correlated to
the number of magnetic features, a long-term variation of the activity level, as it is the case in magnetic cycles, will induce a long-term RV variation.

\vspace{0.5cm}

\noindent
\underline{Target selection}
\vspace{0.1cm}

To study only long-term activity noise, we select stars that are measured more than 3 years, with a minimum of 5 bins of 3 months (each bin must contains at least 3 measurements). In addition, only stars with log(R'$_{HK}$) $< -4.75$ was selected to address solar-like activity. After this selection, we are left with 91 stars out of 451.

\begin{figure}
\begin{center}
\includegraphics[width=11cm]{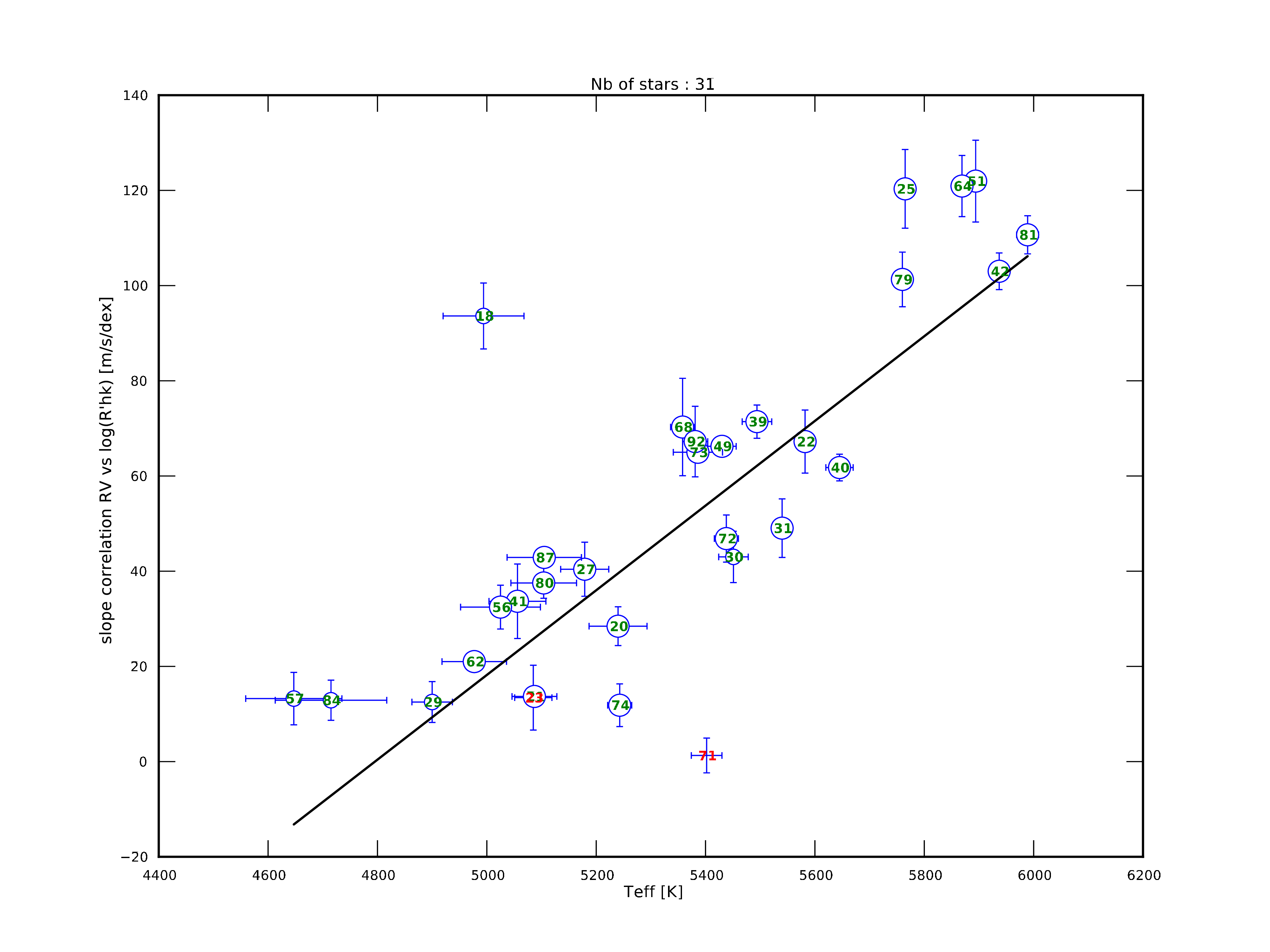}
\caption{Slope of the correlation RV-log(R'$_{HK}$) as a function of the T$_{\textrm{eff}}$. The size of the circle surrounding each star scales with R$_{RV}$, the correlation coefficent between RV and log(R'$_{HK}$). Small size if R$_{RV} > 0.5$ and large size if R$_{RV} > 0.75$. Green numbers correspond to stars which have R$_{FWHM} > 0.5$ or R$_{BIS} > 0.5$ and red numbers (71 and 23), to stars with no well defined correlation.}
\label{fig:2}
\end{center}
\end{figure}

To study potential correlations between the activity index and RV variations (see Dumusque \etal\ (2010), in prep.), we first remove all the planets present in the sample. Then, we select stars in the following way (always for the 3 months binned data):
\begin{itemize}
\item Correlation between the CCF Full Width at Half Maximum (FWHM) and log(R'$_{HK}$), R$_{FWHM} > 0.5$ or between the CCF BISsector span (BIS) and log(R'$_{HK}$), R$_{BIS} > 0.5$,
\item Max(log(R'$_{HK}$)) $> -5$,
\item Peak to peak variation in log(R'$_{HK}$) $> 0.05$.
\end{itemize}
After this second selection, only 31 stars are left. However, a nice correlation appears when we plot these stars in a graph representing the slope of the correlation RV-log(R'$_{HK}$) as a function of the T$_{\textrm{eff}}$ (see Fig \ref{fig:2}). The higher the slope, the more the RV will be affected by magnetic activity. Therefore, RVs of early-Kdwarfs are less affected by magnetic cycles than RVs of early-G dwarfs, making early-K dwarfs optimal targets for very small mass planets surveys. In addition, this relation allow us to correct RVs from the magnetic cycle given the activity index and the T$_{\textrm{eff}}$ of the star.

\begin{figure}
\begin{center}
\includegraphics[width=6cm]{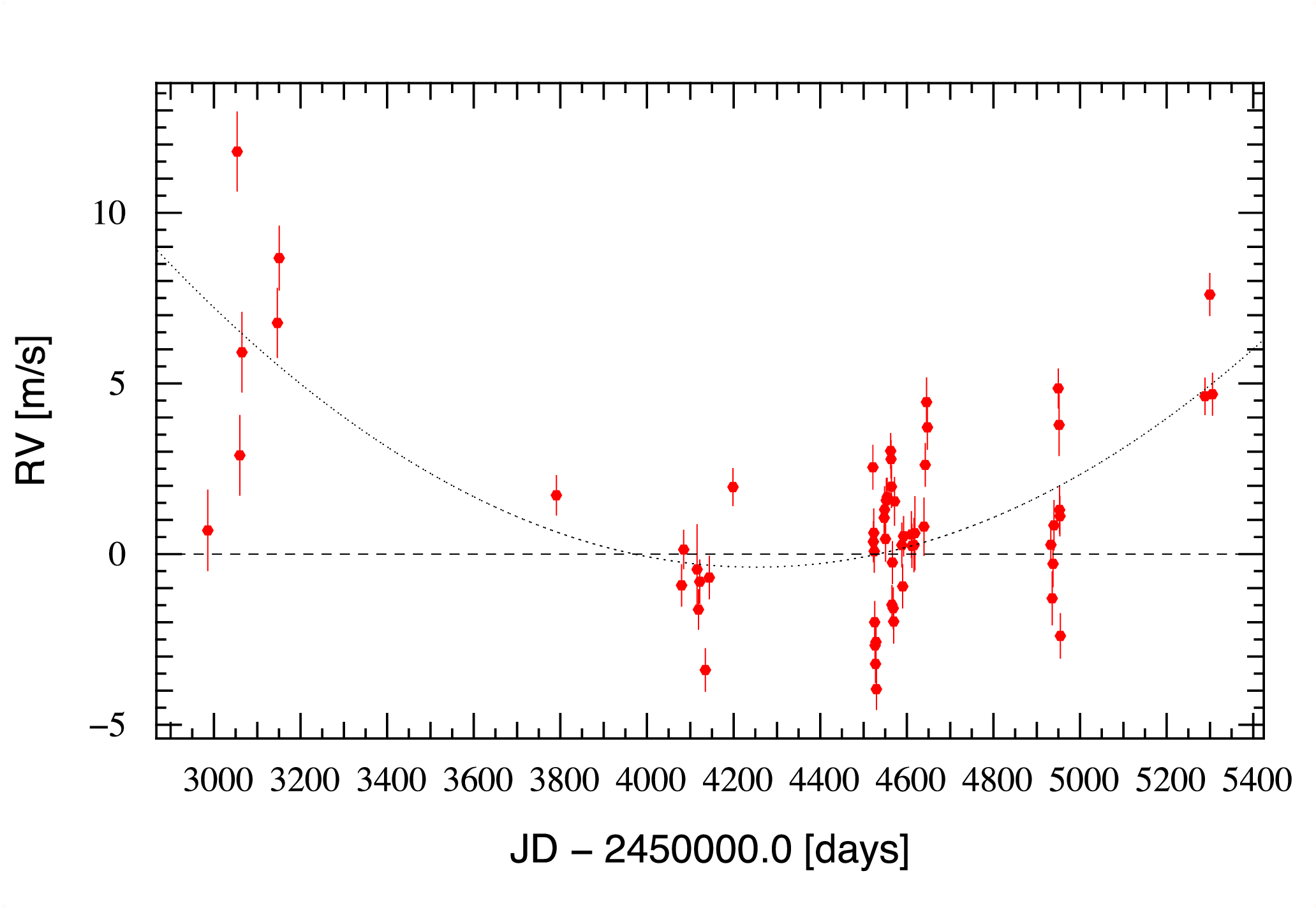}
\includegraphics[width=6cm]{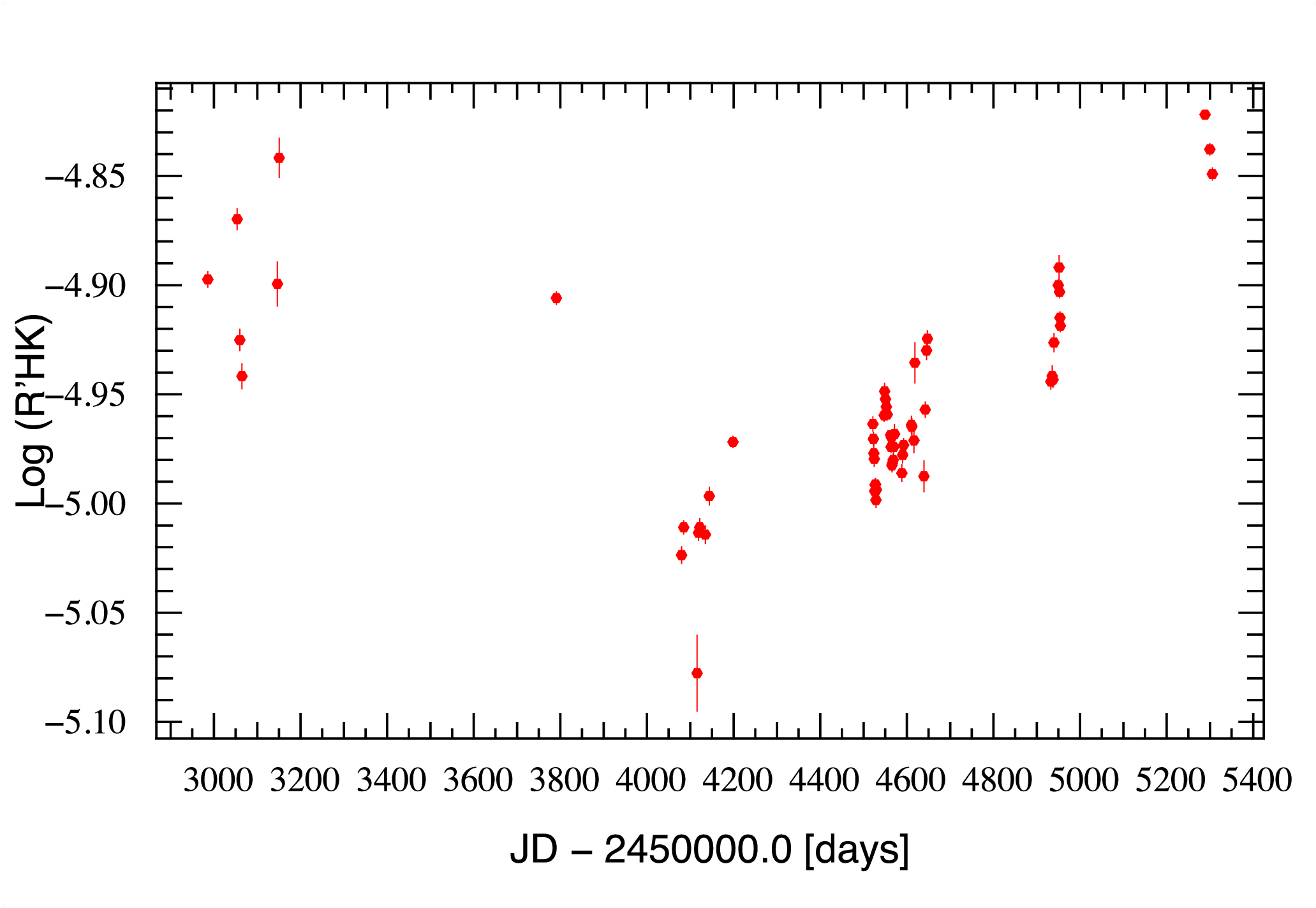}
\caption{Comparison of the long-term RV and activity index (Log(R'HK)) variation. The Pearson correlation between the 2 parameters is very good, for 3 month bins, R = 0.85.}
\label{fig:1}
\end{center}
\end{figure}

\end{document}